\documentclass[
aip, amsmath, amssymb, reprint, floatfix]{revtex4-1}

\usepackage{amsfonts}
\usepackage{amssymb}
\usepackage{array}   
\usepackage{bm}      
\usepackage{braket}
\usepackage{color}
\usepackage{dcolumn} 
\usepackage[T1]{fontenc}
\usepackage{graphicx}
\usepackage{subcaption}   
\usepackage[utf8]{inputenc}
\usepackage{tabularx}
\usepackage{url}
\usepackage[bookmarks=false,colorlinks,citecolor=red]{hyperref}
\usepackage{placeins}   
\usepackage{flafter}  
\usepackage{siunitx}
\usepackage{stackengine}
\usepackage{newtxtext,newtxmath}
\usepackage{ragged2e}
\usepackage{xcolor}

\usepackage{tikz}

\makeatletter
\renewcommand{\fnum@figure}{\textbf{Figure~\thefigure}}
\renewcommand{\fnum@table}{\textbf{Table~\thetable}}
\makeatother


\DeclareUnicodeCharacter{2212}{-}

\newcommand{\BR}[1]{{\color{blue}{#1}}}

\makeatletter
\long\def\@makecaption#1#2{%
    \par
    \vskip\abovecaptionskip
    \begingroup
        \small\rmfamily
        \parindent=0pt
        \leftskip=0pt
        \rightskip=0pt
        \parfillskip=0pt plus 1fil
        \noindent
        \@make@capt@title{#1}{#2}\par
    \endgroup
    \vskip\belowcaptionskip
}
\makeatother

\begin{document}
\IfFileExists{MyCommand.tex}{\include{MyCommand}}{}
\title{Tunable Long-Range Magnetic Exchange in Adatom-Decorated Transition Metal Dichalcogenide Monolayers: An \textit{ab initio} Embedding Model Study}
\author{Xiaohang Jia}
\affiliation{Department of Chemistry, Brown University, Providence, Rhode Island 02912, USA}
\author{Ryan Weiss}
\affiliation{Department of Physics, Brown University, Providence, Rhode Island 02912, USA}
\author{Yusong Bai}
\affiliation{Department of Chemistry, Brown University, Providence, Rhode Island 02912, USA}
\author{Brenda Rubenstein}
\email{Authors to whom correspondence should be addressed: Brenda Rubenstein, brenda\_rubenstein@brown.edu}
\affiliation{Department of Chemistry, Brown University, Providence, Rhode Island 02912, USA}
\affiliation{Department of Physics, Brown University, Providence, Rhode Island 02912, USA}
\affiliation{Data Science Institute, Brown University, Providence, Rhode Island 02912, USA}

\date{November 11, 2026}

\begin{abstract}
Engineering magnetic interactions in two-dimensional materials offers a promising route toward programmable magnetic phases and reconfigurable spintronic functionalities. Here, we investigate magnetic adatoms on metallic monolayer transition-metal dichalcogenides (TMDs) as a platform for tunable indirect exchange, focusing on Fe adatoms on monolayer TaS$_2$. In the intermediate-distance regime, DFT+$U$ calculations reveal atomic-scale oscillations in the exchange coupling between two Fe adatoms and show that biaxial strain tunes the coupling. We develop a first-principles embedding model that reproduces the DFT-calculated pair exchange and establishes a substantial RKKY contribution to the exchange; controlled comparisons further identify RKKY-mediated coupling as the primary origin of the exchange's strain tunability. These results point to combining adatom arrangement with strain as a means of engineering reconfigurable magnetic interactions in metallic monolayer TMDs.
\end{abstract}
\maketitle

\pagenumbering{arabic}

\section{Introduction \label{sec-introduction}}

The strong responsiveness of two-dimensional van der Waals magnets to electrostatic,\cite{jiang2018electric}\cite{eom2023voltage} interfacial,\cite{zhong2017van} and structural perturbations\cite{yao2024moire} has made them promising platforms for electrically-reconfigurable spintronic functionalities.\cite{gong2019two} Their reduced dimensionality and potential for heterostructure engineering have opened access to competing magnetic ground states\cite{xu2022coexisting,zhang20242d} and nontrivial spin textures\cite{song2021direct,birch2022history} that are difficult to stabilize or interrogate in bulk materials. However, deliberate control over 2D magnetism has long been a formidable challenge. While the discovery of intrinsic 2D magnets such as CrI$_3$ and Cr$_2$Ge$_2$Te$_6$ demonstrated that magnetic order can persist in the atomically-thin limit,\cite{huang2017layer,gong2017discovery} the search for robust long-range magnetic order in two dimensions remains a fundamental concern, as highlighted by the Mermin--Wagner theorem.\cite{mermin1966absence}
Further explorations of the tunability of two-dimensional magnetic orders are broadly conducted based on these intrinsic magnetic materials,\cite{song2021direct,xie2023evidence,yao2024moire} where the exchange interaction is dominated by short-range exchange interactions that are strictly determined by the geometries and chemical environments of the magnetic elements, leaving only a comparatively narrow window for \textit{in situ} control.

To unlock a richer degree of tunability in two-dimensional magnetism, it is essential to move beyond the conventional constraints imposed by short-range exchange and fixed lattice geometries, and instead engineer magnetic coupling directly at the electronic structure level by harnessing itinerant charge carriers. The Ruderman--Kittel--Kasuya--Yosida (RKKY) interaction provides such a mechanism, in which localized magnetic moments couple indirectly over long distances through the spin polarization of itinerant electrons in the host.\cite{rkky_k,rkky_rk,rkky_y,rkky_vleck} Unlike superexchange, \cite{anderson1959new,kanamori1959superexchange} which is set primarily by the local bonding geometry, the RKKY interaction is governed by the low-energy electronic structure of the mediator, particularly the Fermi-surface geometry and the associated caliper wave vectors.\cite{roth1966generalization,mirbt1996calculated} In the context of 2D magnetism, this long-range and oscillatory interaction profile is especially appealing, affording a robust structural handle to engineer magnetic order via the spatial configuration of the moment array:\cite{girovsky2017long,khajetoorians2012atom} modulating the separation and arrangement of the moments can directly control the sign and magnitude of the exchange coupling. While this physics has been explored most extensively for magnetic adatoms on bulk metallic systems,\cite{admolecule1,admolecule3,khajetoorians2011realizing} 2D vdW materials offer a more tunable platform for such indirect exchange.\cite{he2013experimental, gorman2013rkky} By virtue of their strong susceptibility to external perturbations, these atomically-thin hosts permit the underlying electronic structure to be tailored on demand, thereby transforming the mediated exchange from a geometry-determined parameter into an \textit{in situ} programmable degree of freedom.

Metallic TMDs, particularly the TaS$_2$ family, provide a natural platform for investigating RKKY exchange in the monolayer limit, owing to their experimental accessibility as monolayers\cite{vavno2023evidence,mehlich2024dislocation} and itinerant carriers capable of mediating long-range indirect exchange.\cite{sanders2016crystalline} Indeed, RKKY-mediated magnetic order has been reported in Fe-intercalated TaS$_2$ compounds in the bulk limit,\cite{ko2011rkky} and more recently, chiral magnetic order has been discussed in intercalated magnetic atoms in TaS$_2$ driven by inversion-symmetry breaking and spin--orbit coupling.\cite{xie2023comparative,camerano2026emergent}  Despite this progress, a systematic understanding of RKKY exchange in the genuine 2D limit---where localized moments are hosted on a metallic monolayer TMD surface and coupled solely through itinerant electrons---is still limited. Moreover, establishing a practical control knob to continuously tune both the magnitude and phase of RKKY exchange remains highly desirable.

In this work, we focus on monolayer TaS$_2$ with Fe adatoms to illustrate a path to tunable two-dimensional magnetic order via RKKY interactions. Density Functional Theory (DFT) calculations reveal an atomic-scale oscillatory exchange between Fe moments mediated by TaS$_2$ as well as this material's high sensitivity to lattice strain, which can efficiently tune its magnitude and sign. To clarify the microscopic origin of the oscillatory exchange beyond the weak-coupling regime of conventional second-order perturbation RKKY, \cite{losada2023nonperturbative,rusin2020exact} we develop a nonperturbative, first-principles-based effective scattering model inspired by the magnetic force theorem.\cite{liechtenstein1984exchange,liechtenstein1987local}
This model reproduces the oscillatory exchange obtained from DFT and corroborates its RKKY origin. Furthermore, it reveals how strain-induced changes in the host electronic structure and Fe--host hybridization modify the RKKY coupling, accounting for the observed tunability of both its magnitude and sign. Altogether, this work provides a recipe for designing -- and studying -- long-range magnetic interactions in 2D materials that can generalize beyond the specific TaS$_2$ example studied here to broad classes of low-dimensional materials.

\section{Methods \label{sec-Methods}}
\subsection{Density Functional Theory (DFT) Calculations \label{sec-DFT}}
Density functional theory (DFT) calculations using a plane-wave basis and projector-augmented-wave (PAW) pseudopotentials \cite{PAW} were carried out using the Vienna \textit{ab initio} simulation package (VASP). \cite{VASP} Perdew–Burke–Ernzerhof (PBE) functionals were adopted to describe the exchange-correlation energy.\cite{PBE} Due to the strong correlation of the $d$ electrons in the transition metal elements, we applied the DFT+$U$ method \cite{DFT+U} with an effective on-site Coulomb parameter of $U = 4$\,eV and an exchange parameter of $J_{\mathrm{Hund}} = 1$\,eV for Fe atoms based on benchmarks from previous literature.\cite{sarkar2022magnetism} The van der Waals interaction was approximated by the DFT-D3 method.\cite{vdW} For all calculations in this work, a 20\,\AA{} vacuum layer was used to eliminate interactions from image atoms.

A $12 \times 12$ supercell of $\text{TaS}_2$ with two identical magnetic adatoms was utilized to investigate the RKKY behavior. A $\Gamma$-centered $k$-point mesh of $1\times 1 \times 1$ was used. The plane-wave energy cutoff was set to 500\,eV. 

The atomic positions for all structures were optimized with a convergence threshold for the maximum force of 0.02\,eV/\AA{}. The collinear magnetic calculations were conducted with an energy convergence threshold set to $10^{-7}$\,eV to capture the energy difference between the antiferromagnetic and ferromagnetic orders of the adatoms. The Ising-like exchange interaction constant $J_{AB}$ was determined by 

\begin{equation}
    J_{AB} = \frac{E_{\mathrm{AFM}} - E_{\mathrm{FM}}}{2},
\end{equation}
where $A$ and $B$ indicate the two adatoms, and $E_{\mathrm{FM}}$ ($E_{\mathrm{AFM}}$) is the energy of the ferromagnetic (antiferromagnetic) configuration. With this convention, $J_{AB}>0$ favors FM alignment and $J_{AB}<0$ favors AFM alignment.

\subsection{Effective Embedding Model \label{sec-Model}}

To describe exchange interactions at short and intermediate separations, we construct a Wannier-based \textit{ab initio} embedding model. Explicit calculations of dilute impurities in very large supercells are computationally costly, so we instead stitch together the pristine-host Hamiltonian $\hat H_0$, obtained from a TaS$_2$ unit cell, and the impurity contribution $\Delta\hat H_{\mathrm{imp}}$, extracted from a $4\times4$ supercell containing a single Fe adatom. We use the same projection-only Wannier orbitals in both calculations so that their Hamiltonians share a consistent gauge.\cite{berlijn2011can}

In this construction, the single-impurity Hamiltonian separates the pristine host from the terms introduced by Fe:

\begin{equation}
\begin{aligned}
    \hat H_{\mathrm{1imp}}
    &=\hat H_0+\Delta\hat H_{\mathrm{imp}}\\
    &=\sum_\sigma\Big[
    \hat{\boldsymbol c}_{h\sigma}^{\dagger}
    H_0^\sigma 
    \hat{\boldsymbol c}_{h\sigma}
    +\hat{\boldsymbol c}_{h\sigma}^{\dagger}
    \Delta H_{hh}^\sigma
    \hat{\boldsymbol c}_{h\sigma}\\
    &\qquad+\hat{\boldsymbol d}_{\sigma}^{\dagger}
    H_{dh}^\sigma\hat{\boldsymbol c}_{h\sigma}
    +
    \hat{\boldsymbol c}_{h\sigma}^{\dagger}
    H_{hd}^\sigma\hat{\boldsymbol d}_{\sigma}\\
    &\qquad+\hat{\boldsymbol d}_{\sigma}^{\dagger}
    H_{dd}^\sigma\hat{\boldsymbol d}_{\sigma}
    \Big].
\end{aligned}
\label{eq:embedded_draft_single_impurity}
\end{equation}
Here, $\hat{\boldsymbol c}_{h\sigma}$ and $\hat{\boldsymbol d}_\sigma$ ($\hat{\boldsymbol c}_{h\sigma}^\dagger$ and $\hat{\boldsymbol d}_\sigma^\dagger$) annihilate (create) electrons with spin $\sigma$ in the host and Fe subspaces, respectively. In our setup, the host subspace contains five Ta $5d$ and three S $3p$ orbitals per atom; the Fe subspace contains its five $3d$ orbitals. The impurity contribution $\Delta\hat H_{\mathrm{imp}}$ is split into three parts: $\Delta H_{hh}^\sigma$ is the change in the host block relative to a frozen, Fe-removed reference, $H_{dd}^\sigma$ describes the Fe orbital energies and spin splitting, and $H_{hd}^\sigma=(H_{dh}^\sigma)^\dagger$ couples the Fe and host subspaces.

By integrating out the d-orbital basis, we produce an embedding model in which the impurities appear as a scattering term, making it convenient to calculate the exchange interaction from the Lloyd formula:
\begin{equation}
    \hat H_{\mathrm{model}}^\sigma(E)
    =\hat H_0^\sigma
    +\sum_{I=1}^{m}
    \hat{\boldsymbol c}_{I\sigma}^{\dagger}
    \widetilde V_I^\sigma(E)
    \hat{\boldsymbol c}_{I\sigma},
    \label{eq:embedded_draft_embedding}
\end{equation}
where $\hat H_0^\sigma$ is the pristine-host Hamiltonian and  $\widetilde V_I^\sigma$ is the effective scattering potential for impurity $I$. The tilde indicates a spatially localized truncation with radius $r_c$. The detailed derivations underpinning this model may be found in Sec.~I of the Supplementary Information.~\cite{supplemental}

\subsection{Calculation of the Exchange Interaction from the Effective Embedding Model}

For two impurities $A$ and $B$, we use the multi-impurity Lloyd formula\cite{lloyd1967wave, oroszlany2019exchange} to calculate their pair interaction without truncating the scattering at second order. Let $\mathcal C_I$ denote the finite set of host Wannier orbitals retained around impurity $I$ (within $r_c$) and let $G_{IJ}^{0,\sigma}$ describe propagation through the pristine host from cluster $\mathcal C_J$ to cluster $\mathcal C_I$. For a chosen collinear spin configuration $\mathcal S$, the pair contribution is obtained by subtracting the scattering contributions of the two isolated impurities from that of the pair:

\begin{align}
&\begin{aligned}
\Omega_{AB}^{\mathcal S}
&\equiv \Delta\Omega^{\mathcal S}[\widetilde V_A+\widetilde V_B]
-\Delta\Omega^{\mathcal S}[\widetilde V_A]
-\Delta\Omega^{\mathcal S}[\widetilde V_B]\\
&=\frac{1}{\pi}\operatorname{Im}\sum_\sigma
\int_{-\infty}^{E_{\mathrm F}}dE\\[-2pt]
&\quad\times\operatorname{Tr}\ln\bigl[
I_{\mathcal C_A}
-T_A^\sigma G_{AB}^{0,\sigma}
T_B^\sigma G_{BA}^{0,\sigma}
\bigr],
\end{aligned}
\label{eq:embedded_draft_lloyd}\\[5pt]
&T_I^\sigma(E)
=\bigl[
I_{\mathcal C_I}
-\widetilde V_I^\sigma(E)G_{II}^{0,\sigma}(E)
\bigr]^{-1}\widetilde V_I^\sigma(E).
\label{eq:embedded_draft_tmatrix}
\end{align}
Here, $E_{\mathrm F}$ is the Fermi level and $\Delta\Omega^{\mathcal S}[\widetilde V]$ is the change in energy caused by the embedded potential or potentials $\widetilde V$. The subtraction removes the two isolated-impurity contributions, leaving the pair interaction. $T_I^\sigma$ is the single-impurity scattering matrix, which sums repeated scattering within $\mathcal C_I$; $I_{\mathcal C_I}$ is the identity matrix in that cluster subspace. The trace is taken over $\mathcal C_A$.

For a collinear spin system, the pairwise exchange coupling is extracted from the pair contributions for the ferromagnetic (FM, parallel moments) and antiferromagnetic (AFM, antiparallel moments) configurations:\begin{equation}
    \begin{aligned}
    J_{AB}
    &=\frac{\Omega_{AB}^{\mathrm{AFM}}-\Omega_{AB}^{\mathrm{FM}}}{2},
    \end{aligned}
    \label{eq:embedded_draft_exchange}
\end{equation}
Details can be found in the Sec.~II of the Supplementary Information~\cite{supplemental} for interested readers.

\section{Results\label{sec-results}}
\subsection{Geometry of Single Fe Adsorption: Lowest Energy Absorption Occurs at the Ta-top Site}

\begin{figure}[htbp]
    \centering
    \includegraphics[width=\columnwidth]{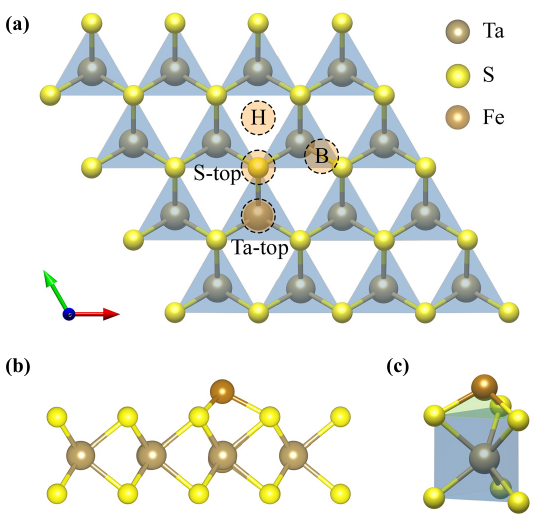}
    \caption{Adsorption geometry and energetics of single Fe adatoms on monolayer $\mathrm{TaS}_2$. (a) High-symmetry adsorption sites: Ta-top, hollow (H), and S-top (S). (b) Top and (c)side views of the optimized Ta-top site adsorption configuration.}
    \label{fig:fig1}
\end{figure}

To determine the energetically-preferred adsorption site for the Fe adatoms on the host TaS$_2$, we performed structural relaxations for Fe localized at the three high-symmetry adsorption sites on monolayer TaS$_2$ [Fig.~\ref{fig:fig1}(a)]: the Ta-top (T), hollow (H), and S-top (S) sites. The calculated adsorption energetics (Table~\ref{tab:fe_adsorption_sites}) show a clear preference for the T site, with the H and S structures lying higher in energy by $\Delta E_{\mathrm{H}}/A = 0.334\,\mathrm{meV}\,\text{\AA}^{-2}$ and $\Delta E_{\mathrm{S}}/A = 0.976\,\mathrm{meV}\,\text{\AA}^{-2}$, respectively. In the optimized Ta-top geometry [Fig.~\ref{fig:fig1}(b,c)], the Fe adatom remains centered above a Ta atom and relaxes to an adsorption height of $h = 2.605\,\text{\AA}$, with Fe--S bond lengths of $2.19\,\text{\AA}$. Given its thermodynamic stability, we adopt this T-site configuration as the structural ground state for all subsequent DFT calculations.

\begin{table}[t]
    \caption{Relative SCF energy densities and adsorption heights of a single Fe adatom at the Ta-top, hollow, and S-top sites of monolayer 1H-TaS$_2$ in a $12\times12$ supercell. Here, $\Delta E=E_{\mathrm{site}}-E_{\mathrm{Ta-top}}$ are calculated from spin-polarized DFT. The heights $h_{\mathrm{top\,S}}$ and $h_{\mathrm{Ta}}$ are measured from the mean planes of the upper S layer and the Ta layer, respectively.}
    \label{tab:fe_adsorption_sites}

    \begin{ruledtabular}
    \begin{tabular}{lccc}
        Adsorption site &
        $\Delta E/A$ (meV/\AA$^2$) &
        $h_{\mathrm{top\,S}}$ (\AA) &
        $h_{\mathrm{Ta}}$ (\AA) \\
        \hline
        Ta-top & 0.000  & 1.041 & 2.605 \\
        Hollow & +0.334 & 0.143 & 1.708 \\
        S-top  & +0.976 & 1.972 & 3.535 \\
    \end{tabular}
    \end{ruledtabular}
\end{table}

\subsection{Magnetic Properties of Single Fe Adsorption: Spin-Density Oscillations Induced by Fe-Ta Local Hybridization}
\begin{figure*}[!t]
    \centering
    \includegraphics[width=\textwidth]{"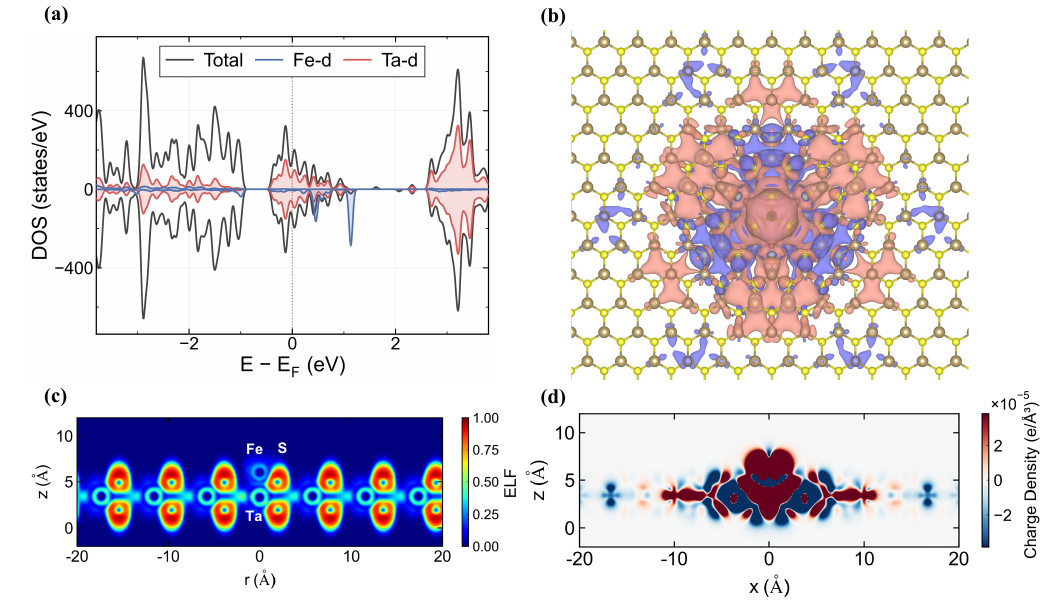"}
    \caption{Electronic and magnetic structures of a single Fe adatom on monolayer $\mathrm{TaS}_2$. (a) Projected density of states (PDOS). The Fe contribution is multiplied by a factor of 20 for visibility. (b) Real-space spin-density distribution showing Friedel-like oscillations. (c) $y$--$z$ cross-sectional view of electron localization function (ELF). (d) $x$--$z$ cross-sectional view of the real-space spin-density distribution.}
    \label{fig:fig2}
\end{figure*}
Before exploring the two-adatom exchange interactions, it is essential to determine the microscopic interactions between the Fe adatom and the $\mathrm{TaS}_2$ host, since the local magnetic moment on Fe, the induced magnetic moments in the $\mathrm{TaS}_2$ substrate, and the local hybridization between the Fe and $\mathrm{TaS}_2$ orbitals are the fundamental factors that dictate the nature of the host-mediated magnetic coupling. In this section, we investigate an isolated Fe adatom in the dilute-impurity limit, modeled by a single Fe located at a T-site of a $12 \times 12$ supercell of monolayer $\mathrm{TaS}_2$, thereby eliminating spurious inter-adatom interactions and isolating the pristine adatom--host response.

    We first evaluate the local magnetic state of the single Fe adatom. The calculated local magnetic moment of the Fe adatom is found to be $3.293\,\mu_{\mathrm{B}}$. This value represents a notable reduction from the $4\,\mu_{\mathrm{B}}$ expected for a high-spin $\mathrm{Fe}$ atom, suggesting a substantial hybridization between the Fe $3d$ states and the itinerant electronic bands of the $\mathrm{TaS}_2$ substrate.\cite{sarkar2022magnetism} Such hybridization is corroborated by the pronounced orbital overlap observed in the projected density of states (PDOS) in Fig.~\ref{fig:fig2}(a). Complementary analysis of the Electron Localization Function (ELF), a real-space descriptor of electron localization and bonding character, provides real-space evidence for this hybridization. Figure~\ref{fig:fig2}(c) shows the ELF in a $y$--$z$ cross section passing through the Fe, S, and Ta atoms, with the coordinate axes defined in Fig.~\ref{fig:fig2}(b). A local ELF enhancement between the Fe adatom and the underlying Ta atom suggests covalent-like bonding and associated local orbital hybridization, consistent with previous findings.\cite{sarkar2022magnetism}

As a consequence of this adatom-host coupling, the spin-density distribution [Fig.~\ref{fig:fig2}(b),(d)] reveals an appreciable, long-range polarization induced by the Fe adatom. In the vicinity of the Fe adatom, the substrate exhibits spatially nonuniform spin polarization, with regions polarized parallel or antiparallel to the Fe moment. The $x$--$z$ cross-sectional view (zigzag direction) in Fig.~\ref{fig:fig2}(d) further shows an alternating sign of the induced spin polarization with increasing distance from the impurity along the $x$ direction, i.e., a Friedel-like oscillatory behavior\cite{yosida1957magnetic}. Such an oscillatory polarization is a hallmark of the itinerant-electron response of a metallic host, implying that the TaS$_2$ surface provides an efficient channel for long-range, carrier-mediated magnetic interactions between adatoms. 

\subsection{Magnetic Exchange Interactions in TaS\textsubscript{2}: RKKY Oscillations in the Short and Intermediate Ranges}

\begin{figure}[htbp]
    \centering
    \includegraphics[width=\columnwidth]{"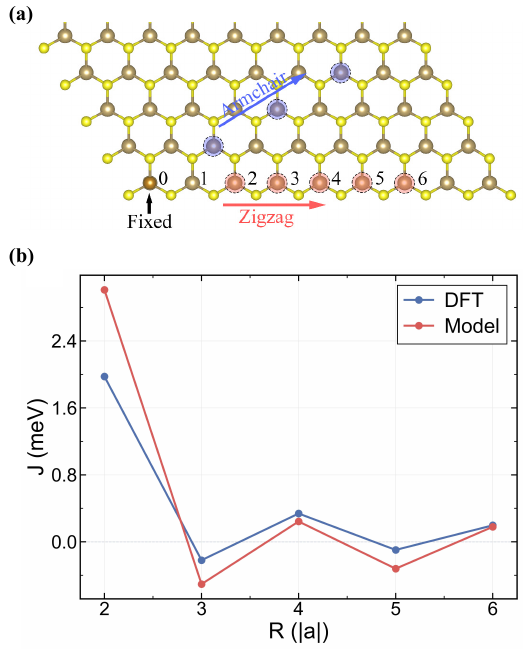"}
    \caption{DFT calculations of two Fe adatoms on monolayer $\mathrm{TaS}_2$: (a) schematic of the DFT configurations; (b) exchange interaction $J(R)$ along the zigzag direction calculated using DFT and the embedding model. The Fe--Fe separation $R$ is expressed in units of the in-plane lattice constant $|a|$.}
    \label{fig:fig3}
\end{figure}

Having modeled the magnetic properties of a single isolated adatom adsorption model, we take a step further to unravel the long-range interaction between two magnetic adatoms. Since we only consider T-site adsorption, it is natural to focus on the zigzag direction to calculate the magnetic exchange interaction. We conducted DFT+$U$ calculations on a $12 \times 12$ supercell of TaS$_2$ containing two Fe adatoms. Methodologically, we fix one Fe adatom at the origin (site 0 in Fig.~\ref{fig:fig3}(a)) and systematically vary the position of the second adatom along the zigzag axis to map the distance-dependent exchange coupling $J(R)$, as illustrated in Fig.~\ref{fig:fig3}(a).

The blue line in Fig.~\ref{fig:fig3}(b) shows the DFT-calculated exchange interactions between the Fe adatoms on the TaS$_2$ supercell. $J(R)$ exhibits an oscillatory behavior with alternating signs as the Fe--Fe separation increases, which is characteristic of an itinerant-electron-mediated indirect exchange in a metallic host.

To identify the origin of this exchange, we calculate the exchange energy using the Lloyd Formula of Eq.~\eqref{eq:embedded_draft_lloyd} within the effective embedding model  Eq.~\eqref{eq:embedded_draft_embedding}, as derived in Sec.~\ref{sec-Methods}. We first consider the same range of Fe--Fe separations as in the DFT calculations. As shown by the red line in Fig.~\ref{fig:fig3}(b), the calculation based on the embedding model agree well with the DFT results, indicating that the effective model preserves the essential physics observed in the DFT calculation.

\begin{figure}[htbp]
    \centering
    \includegraphics[width=\columnwidth]{"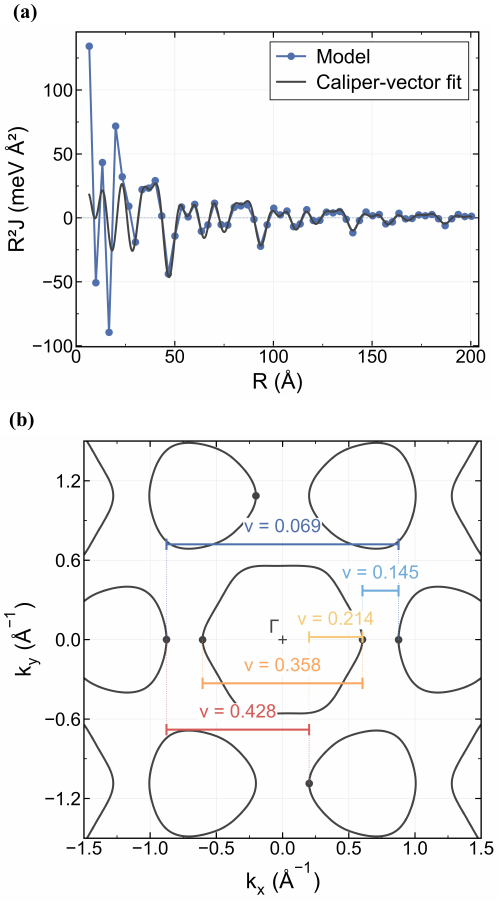"}
    \caption{Long-distance asymptotic behavior of the exchange coupling $J(R)$ and Fermi surface caliper analysis. (a) Long-distance exchange interaction ($50\,\text{\AA} \le R \le 200\,\text{\AA}$) fit by the asymptotic RKKY formula. (b) Caliper vectors of the Fermi surface in the first Brillouin zone.}
    \label{fig:fig4}
\end{figure}

To further test whether the embedding model correctly captures the RKKY contribution, we calculated the exchange interaction in the long-distance limit ($50\,\text{\AA} \le R \le 200\,\text{\AA}$), as denoted by the blue dots in Fig.~\ref{fig:fig4}(a).A Fourier transform of these points extracts a set of irreducible oscillation frequencies, summarized in
Table~\ref{tab:caliper_fft_comparison}. Compared with the irreducible caliper vectors in the first Brillouin zone [Fig.~\ref{fig:fig4}(b)], the oscillating frequencies show a perfect coincidence with the caliper vectors. We subsequently fit the far-field $J(R)$ using the five irreducible caliper vectors in the asymptotic RKKY formula. As shown by the black solid curve in Fig.~\ref{fig:fig4}(a), the RKKY formula accurately matches the model-calculated $J$ for $R \geq 10|\mathbf{a}| \approx 33\,\text{\AA}$. This confirms that the embedding model encapsulates the core physics of the RKKY interaction.

\begin{table}[t]
    \caption{Comparison of the Fermi-surface caliper frequencies ($\nu_{\mathrm{FS}}$)
    with the dominant Fourier frequencies ($\nu_{\mathrm{FFT}}$) of the long-range
    exchange interaction calculated using the embedding model. The related formula may be found in Sec.~III of the Supplementary Information.~\cite{supplemental} }
    \label{tab:caliper_fft_comparison}

    \begin{ruledtabular}
    \begin{tabular}{ccc}
        Channel
        & $\nu_{\mathrm{FS}}$
        & $\nu_{\mathrm{FFT}}$ \\
        \hline
        1 & 0.069 &  0.076  \\ 
        2 & 0.145 &  0.153  \\ 
        3 & 0.214 &  0.211  \\ 
        4 & 0.358 &  0.365  \\ 
        5 & 0.428 &  0.423  \\ 
    \end{tabular}
    \end{ruledtabular}
\end{table}

\begin{figure}[htbp]
    \centering
    \includegraphics[width=\columnwidth]{"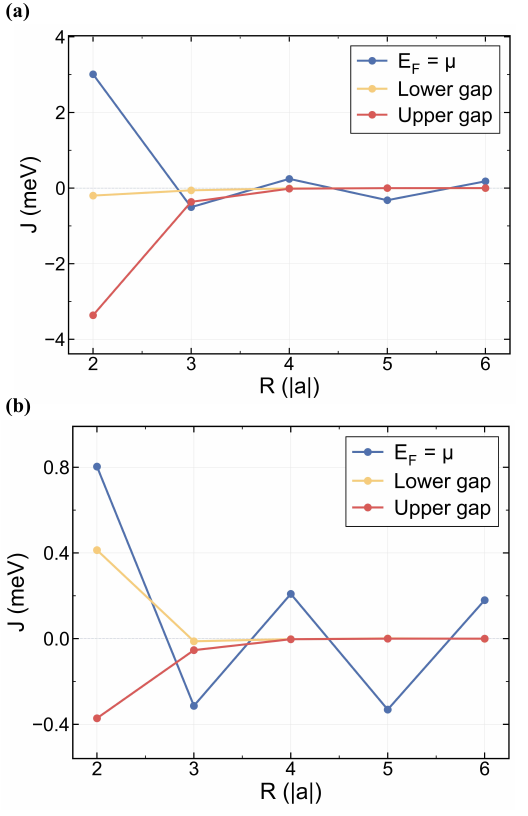"}
    \caption{Exchange interaction $J(R)$ with the Fermi level shifted into the band gaps above and below the active band from an (a) original embedding model calculation and (b) a constrained active band embedding model calculation.}
    \label{fig:fig5}
\end{figure}

With the embedding model validated, we can determine the RKKY contribution to $J$ in the short- and intermediate-range regimes by systematically varying the band filling. Using the original embedding model, we shift the Fermi level into the band gaps above and below the ``active band'' that originally crosses the Fermi level, corresponding to the fully filled and empty active-band limits, respectively. As shown in Fig.~\ref{fig:fig5}(a), in the fully filled case, the $J$ values are comparable to the pristine Fermi energy ($E_F$) for nearest-neighbor distances $n \leq 3$, while the interaction decays rapidly and approaches zero for $n \geq 4$. A similar trend occurs in the empty band case. These results indicate that the exchange at larger separations ($n \geq 4$) highly depends on partial occupation of the active band, consistent with RKKY exchange mediated by itinerant electrons. In contrast, the substantial residual coupling at short separations ($n \leq 3$) in both the fully filled and empty limits suggests additional exchange contributions that do not require partial occupation of this band.

To examine the active-band contribution more directly, we perform a constrained active-band calculation that includes only the Fermi-level-crossing band in the spectral sums for $G_{AB}$ and $G_{BA}$, which describe pristine-host electron propagation between the host-orbital regions surrounding the two Fe adatoms (see Sec.~II of the Supplementary Information~\cite{supplemental}). All other band contributions to these propagators are set to zero, while the local Green's functions and $T$ matrices retain their full-band form. At the original Fermi level, the lattice-scale oscillation is preserved, and the exchange closely reproduces the full-model results for $n\geq4$ [Fig.~\ref{fig:fig5}(b)]. Moving the integration endpoint into the gaps below or above the active band, corresponding to the empty- and filled-band cases, substantially suppresses the exchange, particularly for $n\geq3$, although finite residuals remain at $n=2$. The strong suppression of this exchange in the fully filled and empty limits further establishes its dependence on partial band occupation, supporting a predominantly RKKY-like mechanism.

\subsection{Strain-tunable Magnetic Exchange Interaction: Strain-induced Sign Reversals and Amplitude Modulation of the RKKY Interaction}

We next explore the response of the long-range magnetic interaction to lattice strain. Since RKKY oscillations depend on the host electronic structure, strain-induced changes in the band structure can modify both the magnitude and sign of the exchange. By relating this response to changes in the host electronic structure, we seek further evidence for the RKKY origin of the exchange and assess the possibility of controlling the strength and sign of the magnetic coupling through strain.

In practice, we apply a uniform biaxial strain $\varepsilon$ by scaling the lengths $a$ and $b$ of the two in-plane lattice vectors as
\begin{equation}
a = a_0(1+\varepsilon), \qquad b = b_0(1+\varepsilon),
\end{equation}
while keeping the out-of-plane lattice parameter (vacuum thickness) fixed. In so doing, we keep the original $C_3$ rotational symmetry, allowing us to examine strain-induced changes in the exchange
without introducing additional in-plane symmetry breaking\cite{gorman2013rkky}.

\begin{figure}[htbp]
    \centering
    \includegraphics[width=\columnwidth]{"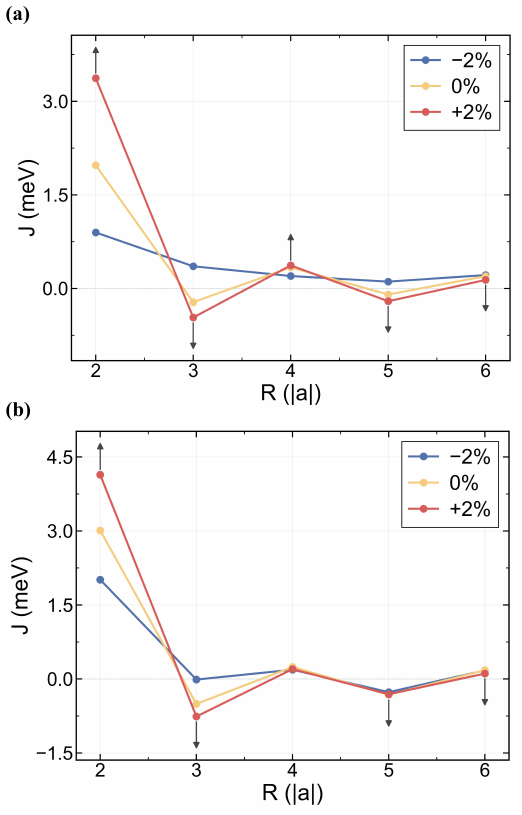"}
    \caption{Exchange interaction $J(R)$ of $\mathrm{TaS}_2$ with different strains calculated by (a) DFT and (b) the embedding model}
    \label{fig:fig6}
\end{figure}

We calculated the magnetic exchange interaction $J(R)$ between two Fe adatoms under uniform biaxial strain $\varepsilon=\{-2\%,-1\%,+1\%,+2\%\}$, where positive (negative) $\varepsilon$ denotes tensile (compressive) strain. As shown in Fig.~\ref{fig:fig6}(a), the RKKY exchange profile $J(R)$ changes smoothly with the applied biaxial strain $\varepsilon$, with strain tuning not only the overall amplitude, but, more importantly, the sign of the interaction. As shown by the blue curve, a moderate compressive strain of $-2\%$ is sufficient to reverse the sign of the exchange at $R=3|a|$ and $5|a|$, switching the preferred coupling from AFM and FM. Such strain-driven sign reversals are difficult to reconcile with a purely short-range superexchange mechanism, and instead support an RKKY-like interpretation. Figure~\ref{fig:fig6}(b) presents the strain response obtained from the embedded model, which captures the overall DFT trends across the investigated range but exhibits weaker strain sensitivity for $n\geq4$. We tentatively attribute this discrepancy to residual errors in stitching the impurity and pristine-host Wannier Hamiltonians together.

To determine the microscopic origin of the strain tunability, we exploit the modular structure of the embedding model to separate the effects of the host electronic structure and the local Fe--host hybridization. Within this framework, the former is encoded primarily in the pristine-host Hamiltonian $H_0(\varepsilon)$, which determines the host Green's function and carrier propagation between the impurities, whereas the latter is contained in the strain-dependent scattering potential $V_{\mathrm{eff}}(\varepsilon)$. We therefore construct two mixed parameterizations,
\begin{equation}
    J_H(\varepsilon)=J[H_0(\varepsilon),V_{\mathrm{eff}}(0)]
\end{equation}
and
\begin{equation}
    J_V(\varepsilon)=J[H_0(0),V_{\mathrm{eff}}(\varepsilon)],
\end{equation}
in which the strain dependence is retained exclusively in the host Hamiltonian or in the impurity potential, respectively. 

\begin{figure}[htbp]
    \centering
    \includegraphics[width=\columnwidth]{"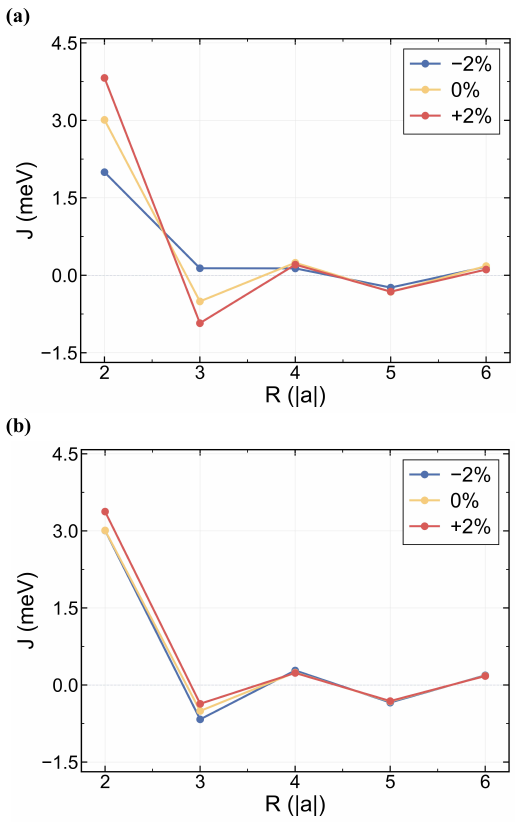"}
    \caption{Exchange interaction $J(R)$ of $\mathrm{TaS}_2$ with different strains calculated from embedding models with (a) fixed $V_{\mathrm{eff}}$ and (b) fixed $H_0$.}
    \label{fig:fig7}
\end{figure}

Fig.~\ref{fig:fig7}(a) shows the result obtained using the strain-dependent $H_0(\varepsilon)$ while keeping $V_{\mathrm{eff}}$ fixed at its unstrained value. This calculation retains the principal strain evolution of the fully parameterized model. Specifically, the exchange at the shorter separations exhibits a pronounced strain dependence, and the $n=3$ coupling changes sign between the compressive and unstrained structures, while the evolution of the remaining couplings is qualitatively consistent with the fully parameterized model. Thus, changes in the host electronic structure alone are sufficient to reproduce the main variations in both the magnitude and sign of the exchange.

The comparative calculation is shown in Fig.~\ref{fig:fig7}(b), where $H_0$ is fixed to its unstrained value and only $V_{\mathrm{eff}}(\varepsilon)$ is varied. In this case, the curves obtained at different strains remain close and preserve essentially the same sign sequence with distance, in contrast to the stronger strain dependence and sign reversals observed in the fully parameterized embedding model and DFT. These results indicate that the strain tunability of the exchange is primarily governed by changes in the host electronic structure that control the RKKY interaction, while changes in the local Fe--host hybridization play a secondary role. Taken together, we conclude that the observed strain tunability arises primarily from changes in the host electronic structure governing the RKKY interaction, and that biaxial strain enables control over both the magnitude and sign of the exchange.

\section{Discussion \label{sec-discussions}}

The present results establish magnetic adatoms on metallic monolayer TMDs as a realistic platform for engineering artificial magnetic orders. A particularly direct route toward their realization is atom-by-atom assembly by scanning probe techniques, as engineered magnetic adatom structures have already been demonstrated on surfaces ranging from metallic substrates to closely related TMDs.\cite{khajetoorians2011realizing, khajetoorians2012atom} In such a setting, the geometry of the adatom array itself becomes a control parameter for magnetic order, as the long-ranged and oscillatory exchange identified here can favor distinct ordered states depending on the spacing and arrangement of the adatoms. At the simplest level, this offers access to collinear ferromagnetic or antiferromagnetic order in one-dimensional chains, while two-dimensional arrays may further support ferromagnetic / antiferromagnetic order, spiral states, or more broadly frustrated phases.

Beyond simple magnetic order, the present system can also provide a platform for noncollinear spin textures. Previous first-principles studies have reported out-of-plane magnetic anisotropy for Fe adatoms on monolayer TMDs,\cite{cong2015enhanced,fang2021spirals} while triangular adatom lattices on TMDs have also been proposed to host noncollinear spin textures.\cite{fang2021spirals} Experimental studies have already shown that spin textures can arise in bulk intercalated TMD compounds.\cite{xie2023comparative,goodge2023multiscale} Although the mechanisms stabilizing noncollinear spin textures may differ between bulk intercalated compounds and monolayers, theoretical studies have predicted bimeron textures in two-dimensional magnets driven by magnetic frustration or Dzyaloshinskii--Moriya interactions, with external magnetic fields providing an additional means of stabilization.\cite{xu2020topological,cha2026first} These considerations point to the broader potential of metallic monolayer TMD adatom platforms for spintronic applications based on controllable nanoscale spin textures.\cite{fang2021spirals,cha2026first}

An additional advantage of this platform is the possibility of controlling magnetic order through strain. As we have shown for the present system, the indirect exchange is highly sensitive to the host electronic structure, so that even moderate strain can modify not only its magnitude but also the preferred magnetic ground state for suitably chosen impurity separations and array geometries. Such a regime appears experimentally accessible, given the maturity of strain engineering in two-dimensional materials and the demonstrated feasibility of percent-level strain in TMD-based platforms. In this sense, ordered impurity arrays could offer a route to strain-controlled switching between distinct magnetic states.

The broader implication is that the behavior identified here is unlikely to be unique to Fe/TaS$_2$. Intercalated metallic TMDs have long provided evidence for itinerant-electron-mediated magnetic interactions across this material family,\cite{wu2022highly, sarkar2022magnetism} suggesting that related monolayer hosts may exhibit similar RKKY-like behavior in the surface-adatom limit. We focus on TaS$_2$ in part because its charge-density-wave order is comparatively weakened in the monolayer limit,\cite{lin2020patterns,sanders2016crystalline} making it a relatively clean platform through which to isolate the role of indirect exchange from competing collective instabilities. Fe/TaS$_2$ should therefore be viewed less as a unique case than as a comparatively clean starting point within a broader class of metallic TMD hosts. Consistent with this picture, our supplementary DFT calculations for Fe/NbS$_2$~\cite{supplemental}  also reveal a similar oscillatory exchange trend, which we attribute to the closely related electronic structures of the NbS$_2$ and TaS$_2$ hosts.

Several limitations of the present model framework should be noted. First, the frozen-potential and pairwise-embedding approximations assume transferable single-impurity perturbations and neglect nonadditive, configuration-dependent changes in the charge redistribution, hybridization, and induced magnetic polarization in the host. Second, using the pristine-host Hamiltonian to describe electron propagation neglects basis mismatches arising from local structural relaxation around the impurity. Differences in orbital centers, shapes, and phases can therefore affect the accuracy of the calculations. Finally, the present implementation is restricted to collinear configurations and isotropic exchange; spin--orbit-induced anisotropic exchange and Dzyaloshinskii--Moriya interactions are not included.

Overall, these results identify magnetic adatoms on metallic monolayer TMDs as a promising setting for tunable indirect exchange in the two-dimensional limit. The combined sensitivity to array geometry and strain suggests a versatile route toward engineered magnetic order and, potentially, noncollinear spin textures on TMD surfaces.

\section{Conclusions \label{sec-conclusions}}
In this work, we have investigated magnetic exchange between Fe adatoms on metallic monolayer TaS$_2$ as a route to engineering tunable magnetic interactions in two dimensions. DFT+$U$ calculations identify the Ta-top site as the preferred adsorption geometry and show substantial hybridization between Fe $3d$ states and itinerant Ta $5d$ bands, together with an oscillatory spin polarization in the host. The pair exchange changes sign with Fe--Fe separation. By combining first-principles Wannier Hamiltonians with an embedding model evaluated using the Lloyd formula, we reproduce the DFT exchange profile and find a substantial RKKY contribution at separations up to $6|\mathbf{a}|$. In the far-field regime, the exchange oscillation frequencies correspond to caliper vectors of the TaS$_2$ Fermi surface. Further, biaxial strain tunes both the magnitude and the sign at selected separations. Calculations from our embedding model show that this strain tunability is dominated by the RKKY interaction. Together, adatom arrangement and strain offer a route to engineering reconfigurable magnetic interactions in metallic monolayer TMDs.

\section{Data Availability}
The data underlying our calculations, fits, visualizations, and other analyses may be found at our project \BR{GitHub page: \url{https://github.com/Neo-626/Wannier-based-embedding-model-for-Fe-TaS-exchange}}. 

\section{AI Disclosure Statement}
AI was used to refine the text of original human-derived drafts of this text.

\section{Acknowledgements \label{sec-ack}}
X.J., R.W., and B.R. were funded by the U.S. Department of Energy through the U.S. Department of Energy, Office of Science, Basic Energy Sciences, Materials Sciences and Engineering Division, as part of the Computational
Materials Sciences Program and Center for Predictive Simulation of Functional Materials. X.J. and B.R. wrote and all authors read, edited, and approved the final manuscript. All authors declare no financial or non-financial competing interests.

\section*{References}
\bibliography{ref}

\end{document}


\IfFileExists{MyCommand.tex}{\include{MyCommand}}{}
\title{Supplementary Information for ``Tunable Long-Range Magnetic Exchange in Adatom-Decorated Transition Metal Dichalcogenide Monolayers: An \textit{ab initio} Embedding Model Study"}
\author{Xiaohang Jia}
\affiliation{Department of Chemistry, Brown University, Providence, Rhode Island 02912, USA}
\author{Ryan Weiss}
\affiliation{Department of Physics, Brown University, Providence, Rhode Island 02912, USA}
\author{Yusong Bai}
\affiliation{Department of Chemistry, Brown University, Providence, Rhode Island 02912, USA}
\author{Brenda Rubenstein}
\email{Authors to whom correspondence should be addressed: Brenda Rubenstein, brenda\_rubenstein@brown.edu}
\affiliation{Department of Chemistry, Brown University, Providence, Rhode Island 02912, USA}

\date{November 11, 2026}

\pacs{}
\maketitle

\section{Single-Impurity Hamiltonian and Downfolding}\label{sec:downfd}

The embedding model is constructed from three DFT reference systems: the pristine TaS$_2$ primitive cell, a relaxed $4\times4$ Fe/TaS$_2$ supercell containing one Fe adatom, and the corresponding Fe-removed supercell. The Fe-removed reference retains the host coordinates of the relaxed Fe/TaS$_2$ structure without further relaxation. The pristine primitive cell provides the host Hamiltonian $H_0^\sigma$, while the two supercell calculations determine the impurity-induced perturbation.

The Wannier Hamiltonians are constructed using Wannier90\cite{pizzi2020wannier90} with projections onto Ta $5d$, S $3p$, and Fe $3d$ orbitals. The primitive cell contains 11 host orbitals, and the single-Fe supercell contains 176 host orbitals and five Fe orbitals per spin channel. We use a projection-only construction without maximal-localization iterations, maintaining consistent orbital ordering and gauge conventions across the reference calculations.~\cite{berlijn2011can} The Wannier sampling meshes are $15\times15\times1$ for the primitive cell and $10\times10\times1$ for the two supercells. The energy windows for the unstrained references are listed in Table~\ref{si:tab-wannier}.

In the host and Fe basis, the single-impurity DFT Hamiltonian takes the block form
\begin{equation}
\begin{aligned}
    \hat H_{\mathrm{1imp}}^{\mathrm{DFT}}
    &=\sum_\sigma
    \hat{\boldsymbol\Psi}_\sigma^\dagger
    \begin{pmatrix}
        H_{hh}^\sigma & H_{hd}^\sigma\\
        H_{dh}^\sigma & H_{dd}^\sigma
    \end{pmatrix}
    \hat{\boldsymbol\Psi}_\sigma,\\
    \hat{\boldsymbol\Psi}_\sigma
    &=\begin{pmatrix}
        \hat{\boldsymbol c}_{h\sigma}\\
        \hat{\boldsymbol d}_\sigma
    \end{pmatrix}.
\end{aligned}
\label{eq:si_single_impurity}
\end{equation}
Here, $\hat{\boldsymbol c}_{h\sigma}$ and $\hat{\boldsymbol d}_\sigma$ collect the host and Fe annihilation operators, respectively. The matrices $H_{hh}^\sigma$ and $H_{dd}^\sigma$ describe electrons in the host and Fe subspaces, while $H_{hd}^\sigma=(H_{dh}^\sigma)^\dagger$ describes their hybridization. The spin dependence of these matrices is obtained from the spin-polarized single-Fe calculation.

After aligning the energy references, the change in the host block is
\begin{equation}
    \Delta H_{hh}^\sigma
    =H_{hh}^\sigma-H_{hh0}^\sigma,
    \label{eq:si_host_difference}
\end{equation}
where $H_{hh0}^\sigma$ is the Hamiltonian of the frozen Fe-removed supercell. Thus, $\Delta H_{hh}^\sigma$ contains the Fe-induced changes in host on-site energies and hoppings at fixed host geometry.

To express the impurity contribution entirely in the host subspace, we integrate out the Fe orbitals. Writing an eigenstate in terms of its host and Fe components, $|\psi_h^\sigma\rangle$ and $|\psi_d^\sigma\rangle$, the Fe block of the eigenvalue equation gives
\begin{equation}
    |\psi_d^\sigma\rangle
    =(EI_d-H_{dd}^\sigma)^{-1}
    H_{dh}^\sigma|\psi_h^\sigma\rangle,
    \label{eq:si_fe_component}
\end{equation}
where $I_d$ is the identity matrix in the five-orbital Fe subspace. Substituting this expression into the host equation and introducing the retarded prescription $E\rightarrow E+i\eta$ yields
\begin{equation}
    H_{\mathrm{host,eff}}^\sigma(E)
    =H_{hh}^\sigma+
    H_{hd}^\sigma
    [(E+i\eta)I_d-H_{dd}^\sigma]^{-1}
    H_{dh}^\sigma,
    \label{eq:si_host_effective}
\end{equation}
where $\eta>0$ is the numerical broadening. The effective scattering potential relative to the Fe-removed reference is therefore
\begin{equation}
\begin{aligned}
    V_{\mathrm{eff}}^\sigma(E)
    &=H_{\mathrm{host,eff}}^\sigma(E)-H_{hh0}^\sigma\\
    &=\Delta H_{hh}^\sigma+
    H_{hd}^\sigma
    [(E+i\eta)I_d-H_{dd}^\sigma]^{-1}
    H_{dh}^\sigma.
\end{aligned}
\label{eq:si_effective_potential}
\end{equation}
The first term describes the modification of the host Hamiltonian, while the second describes a scattering into the Fe orbitals and back into the host. 

\begin{table*}[t]
\centering
\caption{Reference systems and Wannier settings at zero strain. Orbital counts are per
spin channel. Energy windows are the absolute energies used in the
reference calculations, before placing the extracted matrices on a
common energy reference. The Fe-removed calculation uses the same
absolute windows as the single-Fe calculation.}
\label{si:tab-wannier}
\begin{tabular}{lccc}
\hline\hline
 & Pristine primitive cell & Single-Fe $4\times4$ & Fe-removed $4\times4$ \\
\hline
Number of orbitals & 11 & 181 & 176 \\
Wannier $\mathbf k$ mesh & $15\times15\times1$ & $10\times10\times1$ & $10\times10\times1$ \\
Outer window (eV) & $[-10.9506,2.0494]$ & $[-10.5001,2.4999]$ & $[-10.5001,2.4999]$ \\
Frozen window (eV) & $[-6.4506,-3.0506]$ & $[-6.0001,-2.6001]$ & $[-6.0001,-2.6001]$ \\
\hline\hline
\end{tabular}
\end{table*}

We localize this perturbation by projecting onto the host orbitals within a cutoff radius $r_c$:
\begin{equation}
    \widetilde V^\sigma(E)
    =P_{r_c}V_{\mathrm{eff}}^\sigma(E)P_{r_c},
    \label{eq:si_cutoff}
\end{equation}
where $P_{r_c}$ retains all orbitals on host atoms whose three-dimensional positions lie within $r_c$ of the Fe position. Matrix elements involving an orbital outside this region are discarded. We use $r_c=5.3(a/a_0)$~\AA, where $a$ and $a_0$ are the strained and unstrained in-plane lattice constants, respectively. This retains the same shell of 7 Ta and 15 S atoms, corresponding to 80 host orbitals per spin channel. In the following, $\widetilde V^\sigma(E)$ denotes the matrix of this projected potential in the retained orbital basis.

The potential is then translated to each impurity site and embedded in the pristine host:
\begin{equation}
    \hat H_{\mathrm{model}}^\sigma(E)
    =\hat H_0^\sigma+
    \sum_{I=1}^{m}
    \hat{\boldsymbol c}_{I\sigma}^{\dagger}
    \widetilde V_I^\sigma(E)
    \hat{\boldsymbol c}_{I\sigma},
    \label{eq:si_embedding}
\end{equation}
where $\hat{\boldsymbol c}_{I\sigma}$ collects the retained host
orbitals surrounding impurity $I$, and $\hat H_0^\sigma$ is the Hamiltonian of the extended pristine host. Embedding is performed by adding the translated single-impurity scattering potentials to this pristine-host Hamiltonian, with each potential kept fixed.

\section{Projected Green's Function and Lloyd Formula}\label{sec:Lloyd}

The finite region on which $\widetilde V_I^\sigma(E)$ acts defines the cluster subspace
\begin{equation}
    \mathcal C_I
    =\left\{
    |\mathbf R_I+\mathbf L_a,\alpha_a\rangle
    \right\}_{a=1}^{N_{\mathcal C}},
    \label{eq:si_cluster_basis}
\end{equation}
where $\mathbf R_I$ identifies the host cell associated with impurity $I$, $\mathbf L_a$ is a cell offset, and $\alpha_a$ specifies one of the 11 host orbitals within a primitive cell. The same ordered list of cell offsets and orbital indices is used for equivalent impurity sites, with the number of host orbitals $N_{\mathcal C}=80$.

The pristine-host Green's function is calculated from the primitive-cell tight-binding Hamiltonian:
\begin{equation}
    G_{\mathrm{uc}}^{0,\sigma}(\mathbf k,E)
    =\left[(E+i\eta)I_{\mathrm{uc}}
    -H_{\mathrm{uc}}^\sigma(\mathbf k)\right]^{-1},
    \label{eq:si_bloch_green}
\end{equation}
where $I_{\mathrm{uc}}$ is the identity matrix in the 11-orbital primitive-cell basis. Fourier transformation and projection onto the cluster bases give
\begin{equation}
\begin{aligned}
    [G_{IJ}^{0,\sigma}(E)]_{ab}
    &=\langle\mathbf R_I+\mathbf L_a,\alpha_a|
    \hat G^{0,\sigma}(E)
    |\mathbf R_J+\mathbf L_b,\alpha_b\rangle\\
    &=\frac{1}{N_k}\sum_{\mathbf k}
    e^{i\mathbf k\cdot
    (\mathbf R_I-\mathbf R_J+\mathbf L_a-\mathbf L_b)}
    [G_{\mathrm{uc}}^{0,\sigma}(\mathbf k,E)]_{\alpha_a\alpha_b},
\end{aligned}
\label{eq:si_projected_green}
\end{equation}
where $N_k$ is the total number of sampled wave vectors. The Fourier phase follows the convention
$|\mathbf k,\alpha\rangle=N_k^{-1/2}\sum_pe^{i\mathbf k\cdot\mathbf R_p}|\mathbf R_p,\alpha\rangle$.
Thus, $G_{IJ}^{0,\sigma}$ describes propagation from $\mathcal C_J$ to $\mathcal C_I$. In particular, $G_{AB}^{0,\sigma}$ connects the two impurity regions, while $G_{II}^{0,\sigma}$ describes propagation between orbitals within one region through the extended host.

Repeated scattering within each impurity region is summed by the single-impurity $T$ matrix:
\begin{equation}
\begin{aligned}
    T_I^\sigma(E)
    &=\widetilde V_I^\sigma(E)
    +\widetilde V_I^\sigma(E)
    G_{II}^{0,\sigma}(E)T_I^\sigma(E)\\
    &=\left[I_{\mathcal C_I}
    -\widetilde V_I^\sigma(E)G_{II}^{0,\sigma}(E)\right]^{-1}
    \widetilde V_I^\sigma(E),
\end{aligned}
\label{eq:si_tmatrix}
\end{equation}
where $I_{\mathcal C_I}$ is the identity matrix in $\mathcal C_I$.

For two impurities, the pair interaction is obtained by subtracting the isolated-impurity contributions:
\begin{equation}
    \Omega_{AB}^{\mathcal S}
    =\Delta\Omega^{\mathcal S}[\widetilde V_A+\widetilde V_B]
    -\Delta\Omega^{\mathcal S}[\widetilde V_A]
    -\Delta\Omega^{\mathcal S}[\widetilde V_B].
    \label{eq:si_pair_subtraction}
\end{equation}
Here, $\mathcal S$ denotes the FM or AFM configuration, and $\Delta\Omega^{\mathcal S}[V]$ is the scattering contribution to the electronic energy at a common Fermi level $E_{\mathrm F}$. For each strain, $E_{\mathrm F}$ is determined from the pristine-host Hamiltonian at zero temperature with a filling of 6.5 electrons per spin and is kept fixed for all impurity separations and magnetic configurations.

Using the single-impurity $T$ matrices, the pair contribution can be written in the Lloyd form as~\cite{lloyd1967wave,oroszlany2019exchange}
\begin{equation}
    \Omega_{AB}^{\mathcal S}
    =\frac{1}{\pi}\operatorname{Im}\sum_\sigma
    \int_{-\infty}^{E_{\mathrm F}}dE\,
    \operatorname{Tr}\ln\Bigl[
    I_{\mathcal C_A}
    -T_A^\sigma(E)G_{AB}^{0,\sigma}(E)
    T_B^\sigma(E)G_{BA}^{0,\sigma}(E)
    \Bigr].
\label{eq:si_lloyd}
\end{equation}
The trace is taken over $\mathcal C_A$. The product inside the logarithm describes propagation between the two impurity regions and scattering within each region. For the nonmagnetic host without spin--orbit coupling,
$G^{0,\uparrow}=G^{0,\downarrow}$. The FM configuration uses the same spin-channel ordering at both impurities. The AFM configuration is obtained by interchanging
$\widetilde V_B^\uparrow$ and $\widetilde V_B^\downarrow$ while keeping the host Hamiltonian and the single-impurity matrix elements fixed. The exchange coupling is then
\begin{equation}
    J_{AB}
    =\frac{\Omega_{AB}^{\mathrm{AFM}}
    -\Omega_{AB}^{\mathrm{FM}}}{2},
    \label{eq:si_exchange}
\end{equation}
with $J_{AB}>0$ favoring FM alignment and $J_{AB}<0$ favoring AFM alignment.

For separations up to $R=6|a|$, the host Green's functions are evaluated on a $360\times360$ $\Gamma$-centered $\mathbf k$ mesh. The energy integral extends from $E_{\mathrm F}-7.5$~eV to $E_{\mathrm F}$, using an energy spacing of $0.3125$~meV and $\eta=10$~meV. For the far-field calculations up to $R=60|a|$, we use a $360\times360$ $\Gamma$-centered mesh, energy spacing of $0.3125$~meV, and $\eta=10$~meV.

\section{Definition of Irreducible Caliper Vectors}\label{sec:supp-caliper}
The characteristic oscillation periods of the long-range RKKY interaction are determined by caliper vectors of the host Fermi surface \cite{roth1966generalization,mirbt1996calculated}. A caliper vector $\mathbf Q$ connects two Fermi-surface points with parallel local tangents. For RKKY exchange along a given direction, the relevant points have oppositely directed Fermi velocities along that direction. In our calculations, the Fe--Fe separation is $\mathbf R_n=n\mathbf a$, where $\mathbf a$ is the lattice translation along the zigzag direction. The spatial frequency associated with $\mathbf Q$ is therefore
\begin{equation}
    f(\mathbf Q)=\frac{|\mathbf Q\cdot\mathbf a|}{2\pi},
\end{equation}
which measures the number of oscillation cycles per lattice spacing.

Because the exchange is evaluated only at integer multiples of $\mathbf a$, frequencies $f$ and $\pm f+m$, with integer $m$, describe the same sampled oscillation. We remove this redundancy by folding the frequencies into the interval $[0,1/2]$:
\begin{equation}
    \nu(\mathbf Q)=\min_{m\in\mathbb Z}|f(\mathbf Q)-m|.
\end{equation}
Caliper vectors yielding the same $\nu$ contribute to the same oscillation frequency along the sampled direction. We retain one representative for each distinct $\nu$ and refer to these representatives as the irreducible caliper vectors.

\section{Exchange Interaction in Fe/NbS$_2$}\label{sec:supp-fenbs2-exchange}

\begin{figure}[h]
    \centering
    \includegraphics[width=0.8\columnwidth]{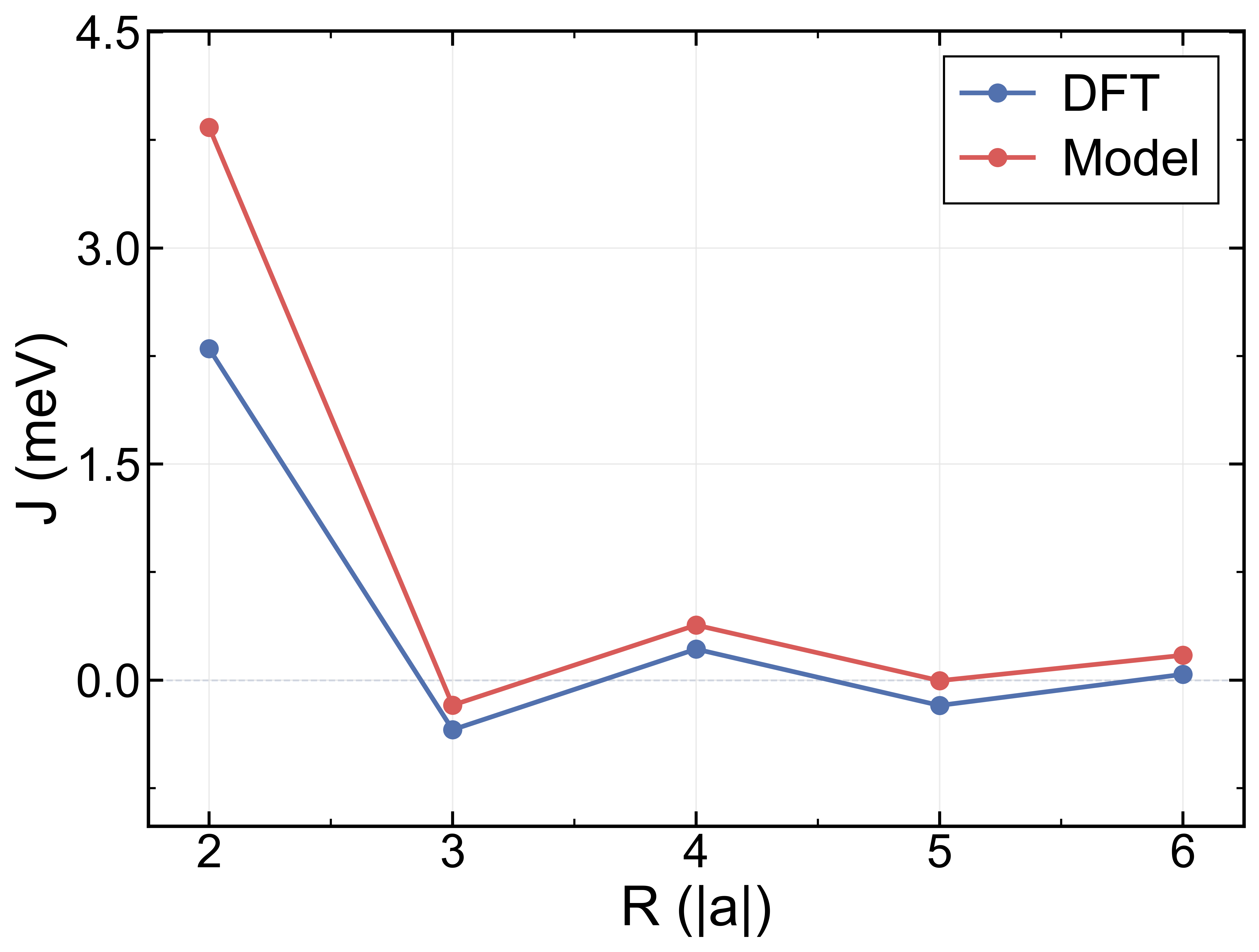}
    \caption{Exchange interaction $J(R)$ along the zigzag direction calculated using DFT and the embedding model.}
    \label{fig:NbS2}
\end{figure}

\clearpage
\bibliography{ref}